\documentclass{cimento_arXiv}
\usepackage{graphicx}  
\usepackage{amsmath}

\title{OPES: On-the-fly Probability Enhanced Sampling Method}
\author{
    Michele Invernizzi\from{inst:eth}\from{inst:usi}\from{inst:iit}
}
\instlist{
  \inst{inst:eth} Department of Physics, ETH Zurich - Lugano, Switzerland
  \inst{inst:usi} Institute of Computational Science,  Universit\`a Svizzera Italiana - Lugano, Switzerland
  \inst{inst:iit} Italian Institute of Technology - Genova, Italy
}

\begin{document}

\maketitle

\begin{abstract}
Molecular simulations are playing an ever increasing role, finding applications in fields as varied as physics, chemistry, biology and material science.
However, many phenomena of interest take place on time scales that are out of reach of standard molecular simulations.
This is known as the sampling problem and over the years several enhanced sampling methods have been developed to mitigate this issue.
We propose a unified approach that puts on the same footing the two most popular families of enhanced sampling methods, and paves the way for novel combined approaches.
The on-the-fly probability enhanced sampling method provides an efficient implementation of such generalized approach, while also focusing on simplicity and robustness.
\end{abstract}

\section{Introduction}
Despite the remarkable improvements over the last decades, both in computational power and in algorithms efficiency, molecular simulations are still limited in their scope by the sampling problem.
Many phenomena of interest, such as protein-ligand binding or crystal nucleation, are rare events that take place on macroscopic time scales and thus would require an impractical amount of computation to be simulated using standard molecular dynamics or Markov-chain Monte Carlo.
To circumvent this problem, a plethora of enhanced sampling methods have been developed that aim at allowing a simulation to visit all the relevant metastable states, without being hindered by kinetic bottlenecks.
Apart from some remarkable exceptions \cite{Bolhuis2002,Skilling2006}, all the most popular enhanced sampling techniques can be roughly grouped in two main families, that we will refer to as collective variables methods and expanded ensembles methods.
We propose a unified perspective on enhanced sampling, that allows us to develop a general method to perform both kinds of sampling in a robust and efficient way.
This new perspective makes enhanced sampling more accessible and simpler to use and can also open up to novel sampling strategies.
The method we developed, called on-the-fly probability enhanced sampling (OPES), is described in detail in Refs.~\cite{Invernizzi2020rethinking} and \cite{Invernizzi2020unified}.
Here we present its main features in a synthetic fashion.

\section{Unified approach}
The goal of enhanced sampling is to increase the probability of observing in a simulation certain rare events, and to do it in such a way that it is still possible to retrieve statistics about the original system.
We call \textit{target distribution} the modified probability distribution that is sampled instead of the Boltzmann one.
Rather than focusing on the various computational techniques used by the different enhanced sampling methods, we propose to group them according to how they define the target distribution they aim to sample.
Following this criteria we can identify two main families.

A first family is the one of methods such as umbrella sampling \cite{Torrie1974} and metadynamics \cite{Laio2002}.
These methods make use of collective variables (CVs) or order parameters that are smooth functions of the atomistic coordinates and encode the slow modes of the system.
In this family, the target distribution is defined by requiring that its marginal probability distribution over such CVs has a given functional form.
Typically the marginal is chosen to be a constant flat distribution, as in adaptive umbrella sampling \cite{Mezei1987}, but other choices are possible, such as the well-tempered distribution often used in metadynamics \cite{Barducci2008}.

A second family includes tempering methods, such as simulated tempering \cite{Marinari1992} and replica exchange \cite{Sugita1999}.
These methods define their target distribution as the combination of slightly different versions of the original system, for example the same system but at higher temperatures.
These target distribution are also known as generalized ensembles or expanded ensembles \cite{Lyubartsev1992}.

The OPES method can be used to sample either kind of target distributions.
It does so by adding to the potential energy of the system $U(\mathbf{x})$ a bias potential $V(\mathbf{x})$ such that the sampled distribution is not the equilibrium Boltzmann distribution, $P(\mathbf{x})\propto e^{-\beta U(\mathbf{x})}$, but the target one, $p^{\text{tg}}(\mathbf{x})$.
This bias potential is defined as
\begin{equation}\label{E:bias}
    V(\mathbf{x})=-\frac{1}{\beta}\log \frac{p^{\text{tg}}(\mathbf{x})}{P(\mathbf{x})}\, ,
\end{equation}
where $\beta$ is the inverse Boltzmann temperature.
The bias potential is not known \textit{a priori}, but it is self-consistently learned during the simulation via an on-the-fly estimate of the probability distributions.
Statistics of the unbiased system can be retrieved via a reweighting procedure, by assuming that the bias is updated in an adiabatic way.
Given any observable $O(\mathbf{x})$, its ensemble average $\langle O\rangle$ over the unbiased system can be estimated via ensemble averages over the sampled biased system:
\begin{equation}\label{E:reweighting}
    \langle O\rangle=\frac{\langle O e^{\beta V}\rangle_V}{\langle e^{\beta V}\rangle_V}\, .
\end{equation}
In this way also free energy differences and free energy surfaces can be estimated \cite{Invernizzi2020rethinking,Invernizzi2020unified}.

\section{OPES for collective variables sampling}\label{S:cv}
Given a set of collective variables $\mathbf{s}=\mathbf{s}(\mathbf{x})$, one can define the marginal probability
\begin{equation}
    P(\mathbf{s})=\int P(\mathbf{x})\,  \delta[\mathbf{s}(\mathbf{x})-\mathbf{s}]\, d \mathbf{x}\, .
\end{equation}
The well-tempered ensemble with respect to these CVs is obtained by requiring that the marginal of the target distribution is $p^{\text{WT}}(\mathbf{s})\propto [P(\mathbf{s})]^{1/\gamma}$, where $\gamma>1$ is know as bias factor.
Notice that the exact target distribution $p^{\text{tg}}(\mathbf{x})$ is not known but this does not constitute a problem. 
In fact, the core requirements are that the corresponding bias potential can be expressed and that the target distribution is easy to sample, thus the kinetic bottlenecks between metastable states are removed.
This is indeed guaranteed for the well-tempered distribution, given that the CVs are chosen properly and the bias factor is large enough.
The case of uniform marginal target distribution can be seen as a special case of the well-tempered one, where $\gamma=\infty$.

When using OPES for CVs sampling we need to estimate $P(\mathbf{s})$.
To do so, we use a weighted kernel density estimation with an automatic kernel merging algorithm, that is explained in detail in Ref.~\cite{Invernizzi2020rethinking}.
We also introduce a regularization term $\epsilon$ and a normalization $Z$ over the explored CV-space.
At step $n$ the bias, Eq.~(\ref{E:bias}), can be written as a function of the CVs:
\begin{equation}\label{E:bias_wt}
    V_n(\mathbf{s})=(1-1/\gamma)\frac{1}{\beta} \log \left( \frac{P_n(\mathbf{s})}{Z_n}+\epsilon \right)\, ,
\end{equation}
where $P_n(\mathbf{s})$ is the estimate of $P(\mathbf{s})$ obtained via reweighting.
Reference \cite{Invernizzi2020rethinking} presents the full derivation of this expression.

\section{OPES for expanded ensembles sampling}\label{S:ee}
To define the expanded ensemble target distribution, we first define a class of probability distributions $P_{\lambda}(\mathbf{x})\propto e^{-\beta U_\lambda(\mathbf{x})}$, where $\lambda$ can be a parameter (e.g.~the temperature) or a set of parameters, and $U_0$ is the original system potential.
For simplicity we only consider nonweighted expanded ensembles, as done in Ref.~\cite{Invernizzi2020unified}.
The expanded ensemble contains a discrete set $\{\lambda\}$ of $N_{\{\lambda\}}$ parameters such that the corresponding $P_\lambda(\mathbf{x})$ have an overlap in the configuration space.
We can write the expanded target distribution as:
\begin{equation}\label{E:target_ee}
    p_{\{\lambda\}}(\mathbf{x})=\frac{1}{N_{\{\lambda\}}}\sum_\lambda P_\lambda(\mathbf{x})\, .
\end{equation}
One can then define the \textit{expansion collective variables} as $\Delta u_\lambda(\mathbf{x})=\beta U_\lambda(\mathbf{x})-\beta U_0(\mathbf{x})$ and use them to write the bias potential at step $n$:
\begin{equation}\label{E:bias_ee}
   V_n(\mathbf{x})=-\frac{1}{\beta}\log \left(\frac{1}{N_{\{\lambda\}}}\sum_\lambda e^{-\Delta u_\lambda(\mathbf{x})+\beta\Delta F_n(\lambda)}\right)\, ,
\end{equation}
where $\Delta F_n(\lambda)$ are the estimates of the free energy differences between the unbiased system $U_0$ and the one at a given $\lambda$.
These are obtained via on-the-fly reweighting, similarly to $P_n(\mathbf{s})$ in Sec.~\ref{S:cv}, but this time without the need for kernel density estimation as $\{\lambda\}$ is a discrete set.
The details of the derivation are explained in Ref.~\cite{Invernizzi2020unified}.

Finally, we notice that often it is possible to rewrite Eq.~(\ref{E:bias_ee}) so that, similarly to Eq.~(\ref{E:bias_wt}), the bias is a function of only a small number of CVs.
For example, in case of a multithermal expanded target distribution the bias can be expressed as a function of the potential energy only \cite{Invernizzi2020unified}.

\section{Example: alanine dipeptide}
\begin{figure}
  \centering
  \includegraphics[width=\columnwidth]{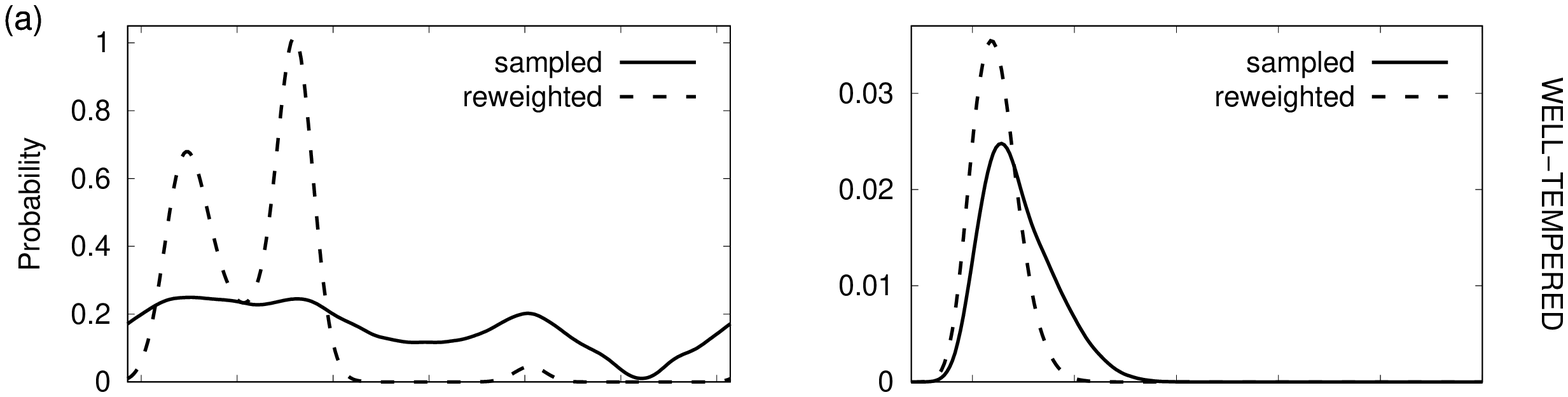}
  \includegraphics[width=\columnwidth]{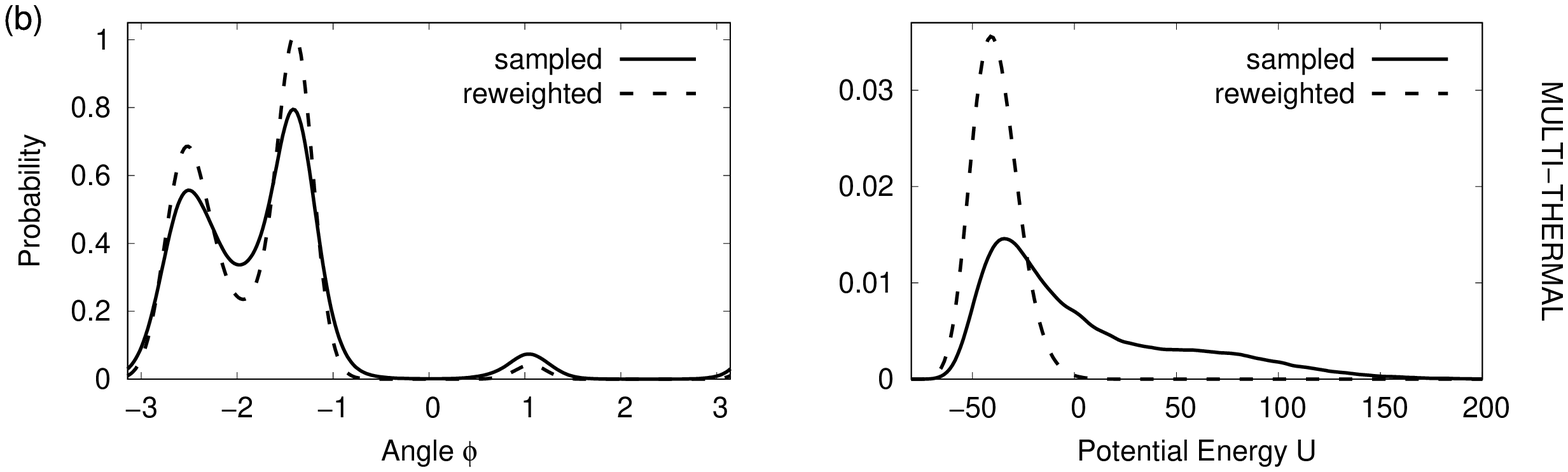}
  \caption{Marginal probabilities over the $\phi$ angle and the potential energy for OPES simulations of alanine dipeptide with different target distributions. 
  The Boltzmann distribution as obtained via reweighting is also shown.
  Top row: well-tempered distribution over $\phi$, $\gamma=50$.
  Bottom row: temperature-expanded ensemble, from 300 to 1000 K.}
  \label{F:ala-wt_mt}
\end{figure}
As an example we consider alanine dipeptide \textit{in vacuum} at 300 K, which is a prototypical system for enhanced sampling.
It presents two main metastable basins, that can be characterized using as CV the torsion angle $\phi$.
The most stable one contains two minima and lays in the region where $\phi$ is negative, while the second basin has one minimum at $\phi\simeq 1$.
A standard unbiased simulation would suffer the sampling problem and almost never make transitions between the two basins.
In Fig.~\ref{F:ala-wt_mt} we show different target distributions that can be used to study alanine, by plotting their marginal along $\phi$ and the potential energy.
On the top, Fig.~\ref{F:ala-wt_mt}a, is the well-tempered ensemble over $\phi$ ($\gamma=50$), which allows the system to easily visit both basins, and greatly increases the probability of sampling intermediate configurations around $\phi=0$.
On the bottom, Fig.~\ref{F:ala-wt_mt}b, is an expanded ensemble that combines four different temperatures, starting from 300 K up to 1000 K.
At higher temperatures alanine explores configurations with higher energy where the barrier between the two basins is smaller.
Also in this case the probability of visiting configurations around $\phi=0$ is increased, but less sensibly.
On the other hand from this second simulation it is possible to retrieve statistics about a whole range of temperatures, instead of 300 K only.
The two target distributions presented are clearly different but we can see that, once reweighting is performed, one retrieves the same underlying Boltzmann probability (dashed lines).

\begin{figure}
  \centering
  \includegraphics[width=\columnwidth]{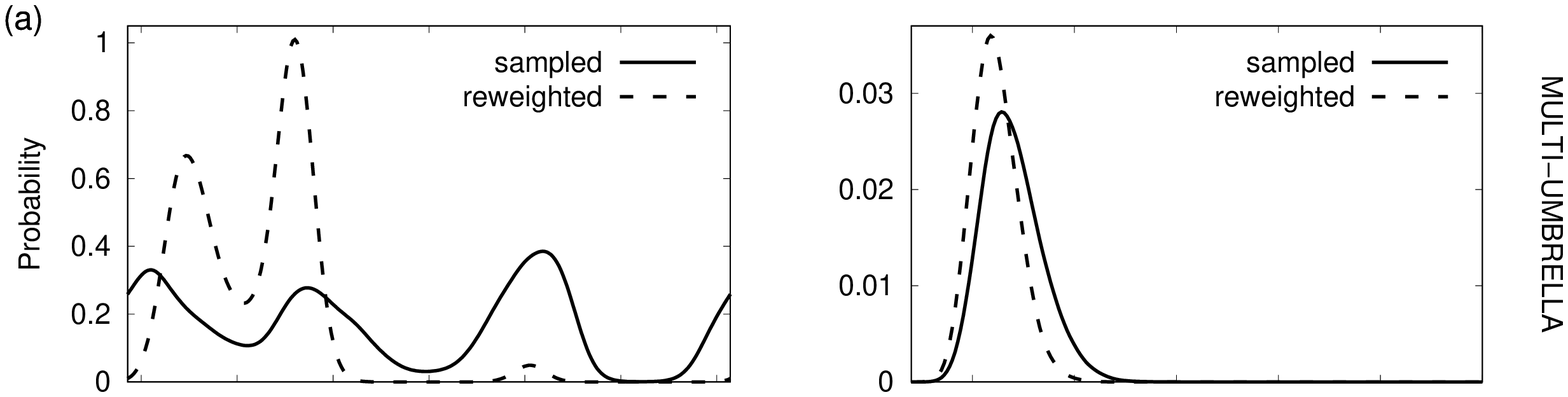}
  \includegraphics[width=\columnwidth]{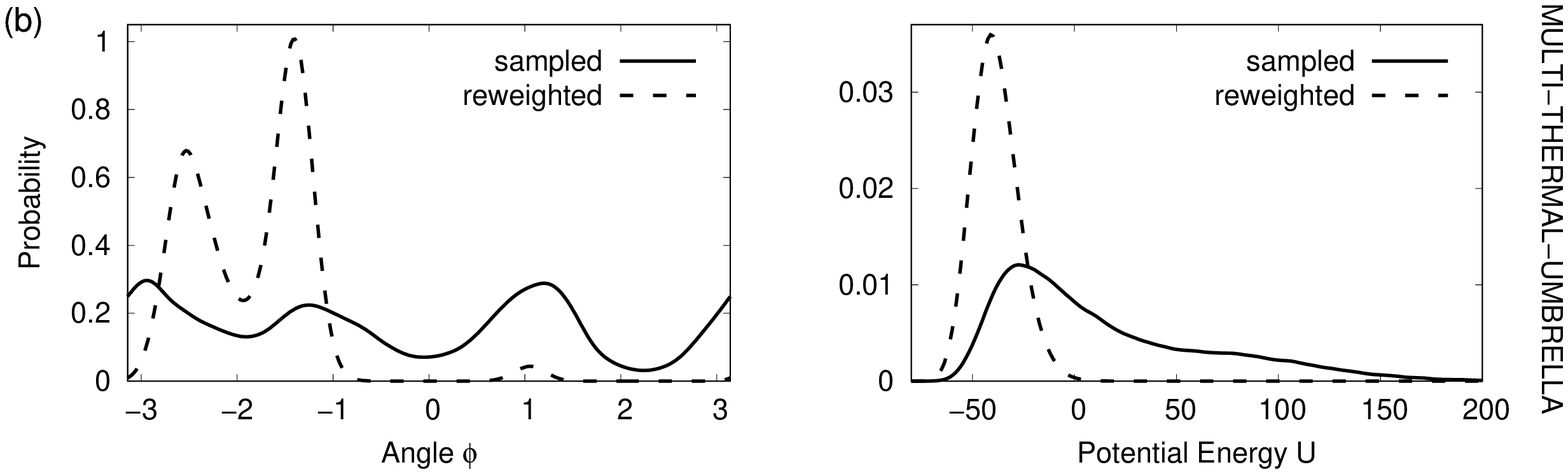}
  \caption{Marginal probabilities over the $\phi$ angle and the potential energy for OPES simulations of alanine dipeptide with different expanded target distributions. 
  The Boltzmann distribution as obtained via reweighting is also shown.
  Top row: multiumbrella distribution over $\phi$, with 43 evenly spaced umbrellas. Notice that it looks quite different from Fig.~\ref{F:ala-wt_mt}a, even though in the limit of infinite umbrellas and $\gamma=\infty$ they would both sample a uniform distribution over $\phi$.
  Bottom row: combined multithermal and multiumbrella distribution, with 43 umbrellas and temperature range from 300 to 1000 K.}
  \label{F:ala-mu_ut}
\end{figure}
It is well known that is possible to enhance the sampling along a CV also using an expanded ensemble as target, e.g.~by combining multiple umbrella sampling distributions \cite{Sugita2000}.
In Fig.~\ref{F:ala-mu_ut}a one can see that a multiumbrella target over $\phi$ can look very different from a well-tempered target over the same CV, Fig.~\ref{F:ala-wt_mt}a.
However, in both cases the system easily transitions from one metastable basin to the other and the probability of being in the transition region (around $\phi=0$) is greatly increased compared to the unbiased case.
By increasing the number of umbrellas in the target we can eventually reach a uniform marginal, similar to the well-tempered one with $\gamma=\infty$.
Finally, it is also possible to combine different expansions in the same target, as shown in Fig.~\ref{F:ala-mu_ut}b, where the used bias is a function both of the angle $\phi$ and of the potential energy $U$.

All the simulation details, together with the inputs and the trajectories are available online on the PLUMED-NEST repository (\texttt{www.plumed-nest.org}, plumID:21.006) \cite{nest}.

\section{Conclusion}
We briefly presented a target distribution perspective on enhanced sampling and the on-the-fly probability enhanced sampling method (OPES), that have been developed in Refs.~\cite{Invernizzi2020rethinking,Invernizzi2020unified}.
OPES is a general and flexible method that can be used to sample different types of target distributions.
It is also easy to use and robust with respect to suboptimal collective variables \cite{Invernizzi2019}.
It has been implemented as a contributed module in the open source library PLUMED \cite{plumed,opes_url}, and has been already used for various applications \cite{Mandelli2020,Bonati2020,Rizzi2021,Piaggi2021,Karmakar2021}.
We believe OPES can be a handy tool for anyone interested in enhanced sampling and it also has the capability of supporting novel types of target distributions.

\acknowledgments
The author thanks Luigi Bonati for carefully reading the manuscript.
This research was supported by the NCCR MARVEL, funded by the Swiss National Science Foundation, and European Union Grant No.~ERC-2014-AdG-670227/VARMET. 


\end{document}